\documentstyle[twoside,fleqn,espcrc2,epsf]{article}

\title{Chiral Symmetry Breaking for Domain Wall Fermions \\
in Quenched Lattice QCD}

\author{
 Lingling Wu\address{Department of Physics, Columbia University,
 New York, NY, 10027, USA }\thanks{This work was done in collaboration
 with T. Blum, P. Chen, N. Christ, C. Christian, C. Dawson, G. Fleming,
 A. Kaehler, X. Liao, G. Liu, C. Malureanu, R. Mawhinney, S. Ohta,
 G. Siegert, A. Soni, C. Sui, P. Vranas, M. Wingate, Y
 Zhestkov(RIKEN-BNL-CU collaboration). 
 This work was supported in part by the Department of Energy and the
 RIKEN BNL Research Center.
 }
}
 
\begin{document}

\def\thepage{CU--TP--991}
\thispagestyle{myheadings}

\begin{abstract}
The domain wall fermion formulation exhibits full chiral
symmetry for finite lattice spacing except for the effects
of mixing between the domain walls.  Close to the continuum
limit these symmetry breaking effects should be described
by a single residual mass.  We determine this mass from
the conservation law obeyed by the conserved axial
current in quenched simulations with $\beta=5.7$ and $6.0$
and domain wall separations varying between $12$ and $48$ on
$8^3 \times 32$ and $16^3 \times 32$ lattices.  Using the 
resulting values for the residual mass we perform two complete 
and independent calculations of the pion decay constant.  Good 
agreement is found between these two methods and with experiment.

\end{abstract}

\maketitle

\section{INTRODUCTION}

An important feature of both domain wall and overlap fermion
formulations is that the chiral limit is disentangled from the
continuum limit. For the case of the domain wall fermions, the chiral
symmetry breaking is solely controlled by the amount of coupling
between the domain walls\cite{kaplan}. 
To exploit the chiral properties of the domain wall fermions, it is
essential to understand these chiral symmetry breaking effects.

We reported earlier\cite{mawhinney,lwu} that for quenched simulations, 
although the quantity $m_\pi^2(m_f=0)$ obtained from the pseudoscalar 
density correlator decreased as the separation between the domain
walls $L_s$ was increased, it did not appear to
vanish in the $L_s \rightarrow \infty$ limit. Further
investigation shows that the chiral limit of the pion mass is distorted
by many factors. Besides the order $a^2$ effects and finite volume effects
common to both dynamical and quenched simulations using domain wall
fermions, the presence of topological near-zero modes in the quenched
approximation significantly affects the determination of the pion
mass\cite{blum2000,matt}. These zero-mode effects can be 
suppressed by going to a larger volume. Other non-linear behaviors
caused by the absence of the fermion determinant in quenched
simulations, such as ``quenched chiral logs'', may also play an important
role in the behavior of the pion mass as the chiral limit is approached. 
These factors make it difficult to determine the degree of chiral
symmetry breaking from examination of 
the pion mass in the quenched chiral limit. The technique 
described in detail below quantifies the chiral symmetry breaking
effects more precisely by a single residual mass measured from the 
extra term in the divergence of the axial current\cite{furman}.

\section{RESIDUAL CHIRAL SYMMETRY BREAKING}

Close to the continuum, the effect of mixing between the domain walls
produces chiral symmetry breaking terms with a coefficient $m_{\rm
res}$, the residual mass, in the low energy effective Lagrangian. 
This results in an effective quark mass of $m_{\rm eff} = m_f + m_{\rm
res}$ entering all low momentum Green's functions. 

\begin{figure}[htb]
\vspace{1pc}
\epsfxsize=\hsize
\epsfbox{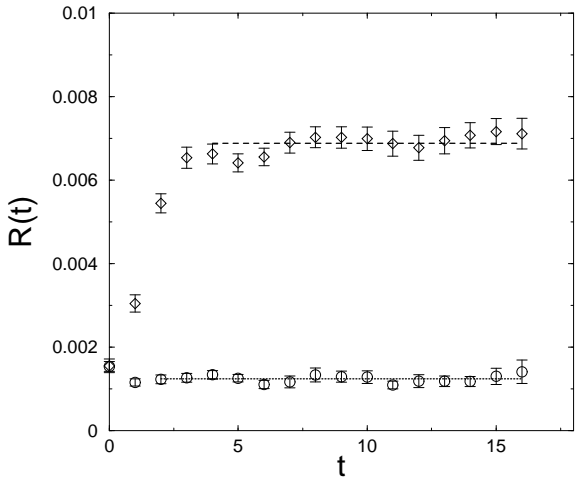}
\caption{$R(t)$ versus $t$ at $\beta=6.0$, $L_s=16$ (circles) and
$\beta=5.7$, $L_s=48$ (diamonds).}
\label{fig:mres_plateau}
\vspace{-1pc}
\end{figure}

We can measure $m_{\rm res}$ starting from the divergence of the axial
current, given by\cite{furman}
\begin{equation}
\Delta_\mu {\cal A}^a_\mu(x) = 2 m_f J^a_5(x) + 2 J^a_{5q}(x).
  \label{eq:axial_cc_diverg}
\end{equation}
Compared with the corresponding continuum expression, 
the additional term from $J^a_{5q}$, referred as the ``mid-point''
contribution, should be equivalent to $m_{\rm res} J^a_5$ up to $O(a^2)$. 
Therefore when t is large enough such that low energy physics dominates,
the ratio
\begin{equation}
  R(t) = \frac{\langle \sum_{\vec{x}} J^a_{5q} (\vec{x}, t)
  \pi^a(0) \rangle } 
  {\langle \sum_{\vec{x}} J^a_5(\vec{x}, t) 
  \pi^a(0) \rangle }
  \label{eq:mres_ratio}
\end{equation}
should be equal to $m_{\rm res}$. Fig. \ref{fig:mres_plateau} shows
that the plateau for $R(t)$ at $\beta=6.0$, $L_s=16$ starts from $t=2$,
and at $\beta=5.7$, $L_s=48$ it starts from $t=4$.

\pagenumbering{arabic}
\addtocounter{page}{1}

\begin{figure}[htb]
\vspace{1pc}
\epsfxsize=\hsize
\epsfbox{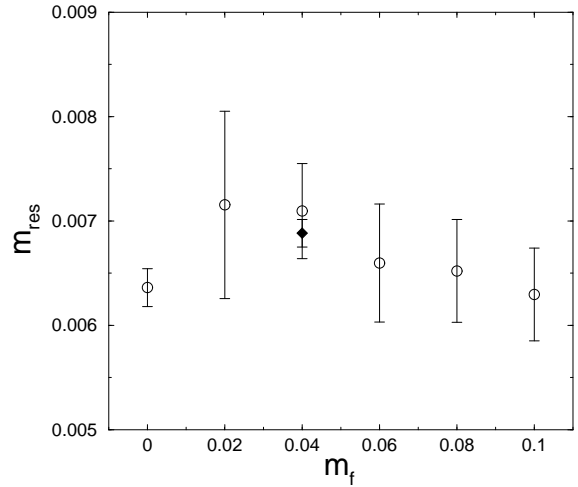}
\caption{The residual mass $m_{\rm res}$ versus $m_f$ at
$\beta=5.70$, $L_s=48$. The open circles are obtained from 106
configurations with volume $16^3 \times 32$, while the closed
diamonds are obtained from 335 configurations with volume $8^3
\times 32$. }
\label{fig:mres_vs_mf}
\vspace{-1pc}
\end{figure}

Being part of the mass term of the low energy effective Lagrangian
close to the continuum, $m_{\rm res}$ should provide a universal
description of the residual chiral symmetry breaking effects 
in all long distance observables. This is demonstrated in Figure
\ref{fig:mres_vs_mf} where $m_{\rm res}$ shows little $m_f$
or volume dependence at $\beta=5.70$, $L_s=48$. 

\begin{figure}[htb]
\vspace{1pc}
\epsfxsize=\hsize
\epsfbox{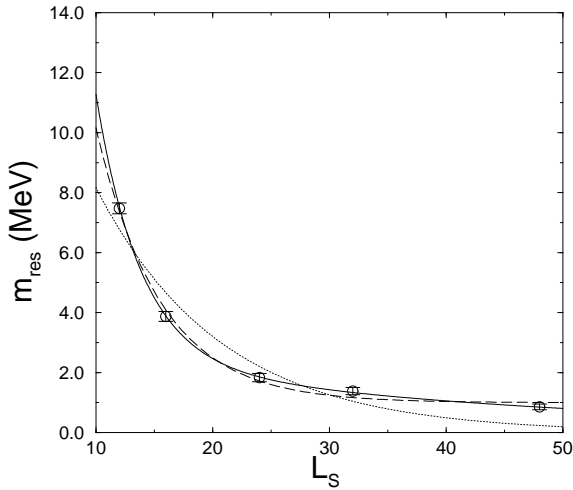}
\caption{The residual mass $m_{\rm res}$ versus $L_s$ at
$\beta=6.0$. The dotted line is a simple exponential fit. The long
dashed line is an exponential plus constant fit. The solid line is a
double-exponential fit.}
\label{fig:mres_vs_ls}
\vspace{-1pc}
\end{figure}

Figure \ref{fig:mres_vs_ls} shows the $L_s$ dependence of $m_{\rm
res}$ at $\beta=6.0$. The residual mass decreases with $L_s$, but the
data points are poorly fit by a simple exponential. If we add a constant to
the function, a much better fit can be obtained. This constant, which
is the residual mass at infinite $L_s$, is about 1 MeV. 
The five points can also be fit extremely well to a 
double-exponential function; this may suggest
the existence of multiple decay modes. The exact asymptotic form for
$m_{\rm res}$ as a function of $L_s$ still needs further
investigation. However for $L_s=16$, $m_{\rm res}$ is already as low
as 3.87(16) MeV in the $\overline{\rm MS}$ scheme at 2 GeV, 
about 1/30 of the strange quark mass. This
value is so small that the residual chiral symmetry breaking is not very
important for these simulations. Even at the stronger coupling $\beta=5.7$
($a^{-1} \sim 1$ GeV), when $L_s$ is increased from 32 to 48, $m_{\rm
res}$ drops from 0.0105(2) to 0.00688(13), which is about 1/14 of the
strange quark mass. Good chiral properties can be obtained by
simulating at this value of $L_s$. 
Studies of this issue can also be found in \cite{CP-PACS1,CP-PACS2}.

\section{CALCULATION OF $f_\pi$}

As a test of these measurements and chiral properties, we calculate the
pion decay constant using the pseudoscalar density correlator, which
requires knowledge of the residual mass, and compare the results
with those obtained from the axial vector current correlator. 

\begin{figure}[htb]
\vspace{1pc}
\epsfxsize=\hsize
\epsfbox{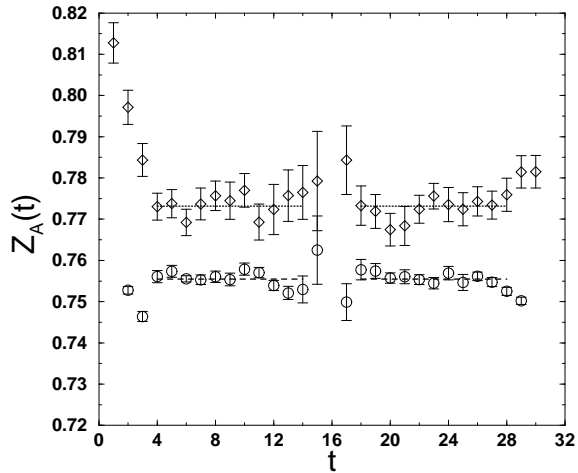}
\caption{$Z_A(t)$ versus $t$ at $\beta=6.0$, $L_s=16$ on $16^3 \times 32$
lattices (circles) and $\beta=5.7$, $L_s=48$ on $8^3 \times 32$
lattices (diamonds).}
\label{fig:za_plateau}
\vspace{-1pc}
\end{figure}

\begin{figure}[htb]
\vspace{1pc}
\epsfxsize=\hsize
\epsfbox{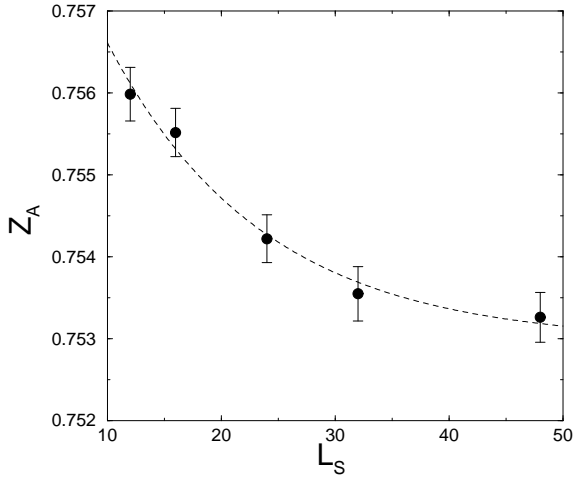}
\caption{$Z_A$ versus $L_s$ at $\beta=6.0$ on $16^3 \times 32$
lattices. The dashed line fits the points to the form of 
$A+B \exp(-\alpha L_s)$. }
\label{fig:za_vs_ls}
\vspace{-1pc}
\end{figure}

Following the definition of the pion decay constant, $f_\pi$ can be
related to the amplitude of the axial current correlator $A_{AA}$:
\begin{equation}
A_{AA} = (\frac{f_\pi}{Z_A})^2 \frac{m_\pi}{2}.
  \label{eq:aa_fpi}
\end{equation}
In Eq. \ref{eq:aa_fpi}, the renormalization factor $Z_A$ is needed to
relate the local axial current to the conserved current in the domain
wall fermion formulation. $Z_A$ can be measured by studying
the large t behavior of the quantity $Z_A(t)$, defined as the ratio of
the coupling between the conserved current and the pion to the coupling 
between the local current and the pion. Figure \ref{fig:za_plateau}
shows that nice plateaus for $Z_A(t)$ can be found over the ranges  
$4 \leq t \leq 14 $ and $18 \leq t \leq 28 $ for both $\beta=6.0$,
$L_s=16$ with $16^3 \times 32$ volume
and $\beta=5.7$, $L_s=48$ with $8^3 \times 32$ volume.
The $L_s$ dependence of $Z_A$ at $\beta=6.0$ is presented in 
Fig. \ref{fig:za_vs_ls} .  For $L_s$ ranging from 12 to 48, the change
in $Z_A$ is less than $1\%$. Therefore, the effect of finite $L_s$ on
the result for $Z_A$ is negligible. 

Another method to calculate $f_\pi$ uses the relation
between $f_\pi$ and the amplitude of the pseudoscalar
correlator $A_{PP}$: 
\begin{equation}
A_{PP} = -(\frac{f_\pi}{m_f+m_{\rm res}})^2 \frac{m_\pi^3}{8},
  \label{eq:pp_fpi}
\end{equation}
which directly depends upon the residual mass $m_{\rm res}$. 

Results for $\beta=5.7$, $L_s=48$ with lattice volume $8^3 \times 32$
show a discrepancy in the $m_f$ dependence of $f_\pi$ between these two
methods. While the difference appears to become somewhat smaller at the
larger volume of $16^3 \times 32$, we
attribute the discrepancy between these two methods of calculating
$f_\pi$ to $O(a^2)$ errors.

\begin{figure}[htb]
\vspace{1pc}
\epsfxsize=\hsize
\epsfbox{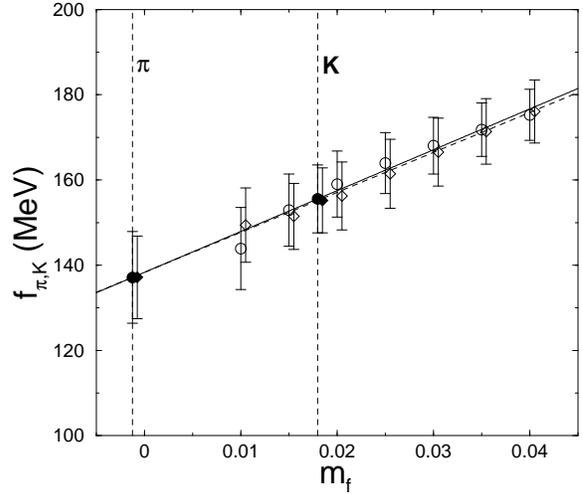}
\caption{$f_{\pi,K}$ versus $m_f$ at $\beta=6.0$ on $16^3 \times 32$
lattices from $\langle A_0^a(x) A_0^a(0) \rangle$ correlator ($\circ$) 
and $\langle \pi^a(x) \pi^a(0) \rangle$ correlator ($\diamond$). The
solid symbols represent the extrapolations to $m_f = - m_{\rm res}$
and interpolations to the kaon mass. The diamonds are slightly 
shifted to the right for clarity.}
\label{fig:fpi_b6_0_2parm}
\vspace{-1pc}
\end{figure}

Figure \ref{fig:fpi_b6_0_2parm} shows the comparison of these two
independent calculations at the weaker coupling $\beta=6.0$. 
Consistent results are seen at each valence quark mass, and the $f_\pi$
values measured by extrapolation to $m_f = -m_{\rm res}$ agree well
with the physical value of 130 MeV, where $m_\rho$ is used to set the
scale. It is important to note that these good properties disappear
if the value for $m_{\rm res}$ in Eq. \ref{eq:pp_fpi}
is replaced by 0 or by the $x$ intercept of the simple linear fit to
$m_\pi^2$. Therefore, the good agreement seen between the two
approaches serves as a consistency check of the residual mass analysis
and a demonstration of the good chiral properties of the domain wall
fermion formulation. 

\section{CONCLUSIONS}

The chiral symmetry breaking effects in the domain wall fermion
formulation can be characterized by a residual mass, which enters
the effective quark mass in the low energy effective Lagrangian. This
quantity can be determined accurately from the additional term in the
divergence of axial current. Good chiral properties can be achieved
for the quenched domain wall formulation for lattice spacing
$a^{-1} \sim 1-2$ GeV. The results for $m_{\rm res}$ are further
checked by our determinations of the pion decay constant.

\section{ACKNOWLEDGMENT}

These calculations were performed on the QCDSP machines at Columbia
and the RIKEN BNL Research Center.

\end{document}